\documentclass[runningheads]{llncs}
\usepackage[T1]{fontenc}
\usepackage{graphicx}
\begin{document}
\title{Towards an LLM-powered\\Social Digital Twinning Platform}
\titlerunning{Towards an LLM-powered Social Digital Twinning Platform}
% If the paper title is too long for the running head, you can set
% an abbreviated paper title here
%
\author{\"{O}nder G\"{u}rcan\orcidID{0000-0001-6982-5658}
\and Vanja Falck\orcidID{0000-0003-0855-5943}
\and Markus G. Rousseau\orcidID{0009-0006-2615-8166}
\and Larissa L. Lima\orcidID{0000-0003-2535-828X} 
%\and Patrycja Antosz\orcidID{0000-0001-6330-1597} (waiting for her contributions)
}
\authorrunning{\"{O}. G\"{u}rcan et al.}
% First names are abbreviated in the running head.
% If there are more than two authors, 'et al.' is used.
%
\institute{
    Center for Modeling Social Systems\\
    NORCE Norwegian Research Center AS\\
    Kristiansand, Norway
}
\maketitle              % typeset the header of the contribution
\begin{abstract}
We present Social Digital Twinner, an innovative social simulation tool for exploring plausible effects of what-if scenarios in complex adaptive social systems. 
The architecture is composed of three seamlessly integrated parts: a data infrastructure featuring real-world data and a multi-dimensionally representative synthetic population of citizens, an LLM-enabled agent-based simulation engine, and a user interface that enable intuitive, natural language interactions with the simulation engine and the artificial agents (i.e. citizens). 
Social Digital Twinner facilitates real-time engagement and empowers stakeholders to collaboratively design, test, and refine intervention measures. 
The approach is promoting a data-driven and evidence-based approach to societal problem-solving. 
We demonstrate the tool’s interactive capabilities by addressing the critical issue of youth school dropouts in Kragerø, Norway, showcasing its ability to create and execute a dedicated social digital twin using natural language. 

\end{abstract}
\section{Introduction}
We define Social Digital Twins (SDTs) as virtual models that faithfully replicate real-world social systems, communities, or societal processes in order to allow decision-makers to analyze and predict complex social phenomena.
SDTs use \textit{extensive} real-world data and recreates the behaviors and interactions of individuals or groups within a society using agent-based modeling and simulation (ABMS). 
By modeling and visualizing the intricate interplay of individual behaviors, collective dynamics, and institutional policies, SDTs enable policymakers, researchers, and stakeholders to test "what-if" scenarios, mitigating the risks of real-world policy experimentation, and to  co-develop solutions to critical societal challenges, including education, employment, and public health.
%This approach facilitates evidence-based decision-making, accelerates policy development, and fosters a participatory environment where stakeholders
%
%By leveraging SDT technology, cities can conduct various social experiments and simulations at the city level, formulating measures to solve diverse and complex issues. 
%This approach supports the realization of safe and sustainable next-generation smart cities by promoting measures to mitigate pandemics, ensure efficient allocation of medical resources, and drive economic growth \footnote{https://www.fujitsu.com/global/about/resources/news/press-releases/2022/0208-01.html, last access on 16/12/2024.}.
%Unlike traditional social simulation systems that often focus on general or abstract agent behavior, a Social Digital Twin (SDT) is grounded in real-world data and operates as a continuously updated, high-fidelity mirror of societal systems. %The term "twin" emphasizes not just simulation but alignment with the actual state of a target population or environment. This alignment enables SDTs to be more immediately actionable and context-specific than generic agent-based models.
%As noted in [4], societal twins aim for ongoing synchronization and feedback with the physical world, distinguishing them from static or scenario-limited simulations.

In recent years, the rising interest in SDTs has enabled the creation of advanced tools to explore and address the complexities of urban dynamics~\cite{Deng2021,Lei2023}. 
Most research on SDT applications focuses on their implementation in the smart city domain~\cite{Chircu2023}.
Examples include ~\cite{Yossef2023,Dembski2020,Schrotter2020,White2021,Cherian2023}.
Yossef Ravid and Aharon-Gutman ~\cite{Yossef2023} incorporated social aspects into the decision-making process for the Hadar neighborhood in Haifa, Israel.
Dembski et al.~\cite{Dembski2020} used social data and volunteered geographic information for the city of Herrenberg in Germany.
Schrotter and Hürzeler \cite{Schrotter2020} developed a 3D spatial data model using Zurich's infrastructure that can be used to test different planning scenarios.
White et al. \cite{White2021} developed a digital twin of the Docklands area in Dublin that allows user interaction and feedback on urban planning and additional simulations on flooding and crowds. 
Cherian et al.~\cite{Cherian2023} developed an agent-based modeling (ABM) framework to address COVID-19 that can simulate the Indian population at multiple scales.
Batty and Milton~\cite{Batty2023} developed a DT for British cities that covers over 8000 urban places.

Despite these advances, most existing SDT solutions remain tightly focused on specific problems rather than offering a flexible, generic platform, as mentioned by Birks et al. ~\cite{Birks2020}. 
Moreover, they often require advanced programming knowledge to set up, customize, or improve, which creates a significant barrier for decision-makers, researchers, or community stakeholders who lack deep technical expertise.
Based on this observation, we present a proof-of-concept system that harnesses the power of Large Language Models (LLMs) to enable natural language-based interactions between artificial agents, the simulation engine, and human users. 
By integrating LLM-driven natural language processing, users can guide and influence the simulation process, co-creating and testing intervention strategies in real time. 
This system not only enhances user engagement but also fosters a participatory approach to policy development and societal problem-solving.

%In this paper, we showcase the tool’s capabilities by addressing the critical issue of youth school drop-out in Kragerø, Norway, demonstrating its potential for dynamic policy experimentation and informed decision-making.
%Our model integrates social and geospatial data and a synthetic population to simulate and analyze the factors contributing to dropout rates. 
%The main goal is to enable stakeholders to identify underlying causes and implement targeted interventions. 
%They can interact with local residents and ask questions. 
%This approach demonstrates the potential of digital twins to provide actionable insights for tackling complex social challenges at the community level.

The contributions of this study are as follows:

\begin{itemize}
    \item We designed a novel platform to create and study dedicated faithful social digital twins using real data and realistic synthetic data.
    \item We exploit the richness of the existing European standardised living condition data (EU-SILC) to enrich the attributes of artificial citizens in social simulations.
    \item We designed an innovative simulation engine that integrates LLMs to enable dynamic, natural language interactions between users, autonomous agents, and the simulation environment.
    \item We showcase the platform's interaction capabilities by using a real societal use case.
\end{itemize}

The organization of this paper is as follows: In Section \ref{sec:Background}, an overview of the social digital twinning approaches is provided. 
Section \ref{sec:Social-Digital-Twin-of-Norway} introduces the proposed social digital twinning framework, focusing on artificial population generation, and roles and actions for integrating LLMs into MAS.
Section \ref{sec:Case-Study-School-Dropout-of-NEETs} demonstrates the tool’s interactive capabilities by addressing the critical issue of youth school dropouts in Kragerø.
Section \ref{sec:Discussion} discusses the implications and challenges of the study. 
Section \ref{sec:Conclusion-and-Future-Work} concludes the paper by summarizing the findings, and outlines potential future work.

%%%%%%%%%%%%%%%%%%%%%%%%%%%%%%%%%%%%%%%%%%%%%%%%%%%%%%%%%%%%%%%%%%%%%%%%

\section{Background and Motivation}
\label{sec:Background}

Digital twins refer to virtual replicas of physical systems or processes that run in parallel with real-world counterparts, often synchronizing in real time. The concept was originally conceived for complex industrial and aerospace applications~\cite{Wang2024}, but it has since expanded across domains ranging from manufacturing and engineering to education and medicine~\cite{Belbachir2025,Akhmedov2023,Laubenbacher2022}. By providing high-fidelity, up-to-date representations of assets and processes, digital twins enable more informed and efficient decision-making in these fields. 

One domain where digital twins have gained significant traction is urban planning. City-scale Urban Digital Twins have emerged as platforms that bridge physical and virtual urban environments~\cite{Ferre2022,Deren2021,Lehtola2022}. For example, projects like Virtual Singapore and Rotterdam’s city twin create comprehensive 3D models of cities integrated with real-time data to support planning and governance. These systems facilitate scenario testing (e.g., evaluating new infrastructure or policies) by allowing interactive exploration of urban dynamics. They also promote stakeholder engagement through shared, current information, thus improving decision-making across different spatial and temporal scales~\cite{Lehtola2022,Lei2023}. On a broader spectrum, national and community-level initiatives such as New Zealand’s ALMA project and MIT’s CityScope Andorra show the growing ambition to simulate societal processes at multiple scales in an interactive manner. 

Social Digital Twins (SDTs) build on the digital twin concept by adding a social layer—modeling individuals, communities, and their interactions, not just physical infrastructure. Traditional smart city twins often overlook human behavior and social processes, creating blind spots in policy analysis and risks to equity and inclusion~\cite{Birks2020}. To address this, researchers advocate for human-centric twins that explicitly capture social dynamics~\cite{Birks2020,Batty2024}. Agent-based modeling (ABM) offers a promising method, representing citizens and organizations as agents whose actions and interactions can be simulated~\cite{Birks2020}. Early examples include CityScope Andorra’s tangible ABM for urban design and the ALMA initiative, which models a national population, showing SDTs' potential for addressing complex societal challenges.

Despite recent progress, most existing SDT solutions are still bespoke and siloed, requiring significant technical expertise to develop or modify~\cite{Birks2020}. 
This hinders flexibility and wider adoption. 
To overcome this, we propose using Large Language Models (LLMs) to improve the accessibility, realism, and adaptability of SDT platforms. 
LLMs can act as both intuitive interfaces and generative engines for agent behavior. 
This allows stakeholders to interact via natural language—lowering the barrier for non-experts—and enables agents to produce lifelike, context-aware responses \cite{Gurcan2024}. 
LLM-driven agents can also be easily adapted to new scenarios using language prompts, eliminating the need for reprogramming. In short, integrating LLMs into SDTs can democratize their use and enable richer, more flexible simulations.

\section{Social Digital Twinner}
\label{sec:Social-Digital-Twin-of-Norway}

\begin{figure}[h]
  \centering
  \includegraphics[width=0.99\linewidth]{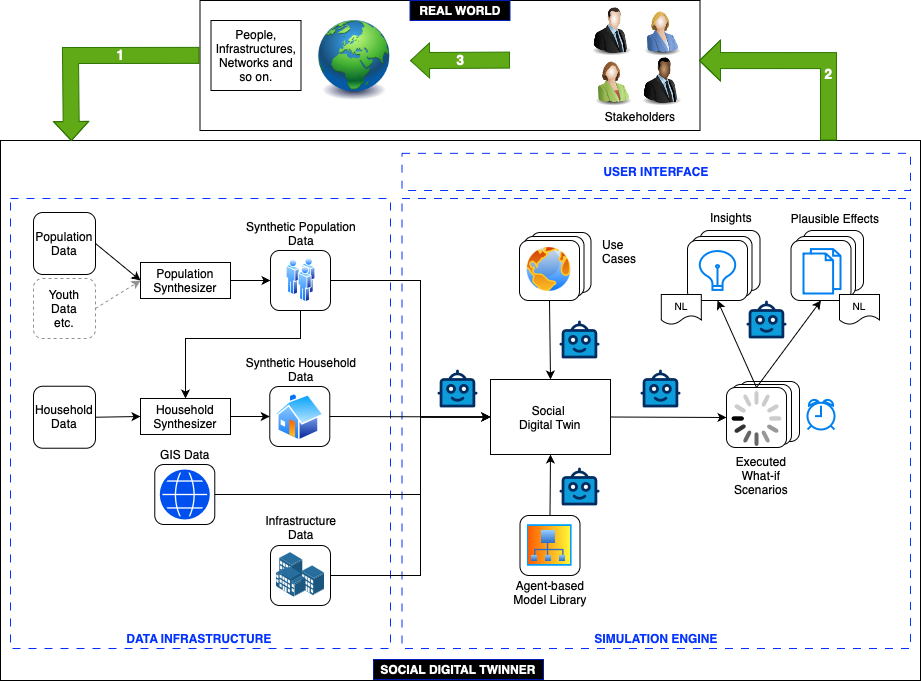}
  \caption{Architectural Framework of the LLM-enabled Social Digital Twinner}
  \label{fig:SDT_Architecture}
\end{figure}

Social Digital Twinner (shortly Twinner) is a tool that aims to produce cutting-edge research on social and societal challenges using agent-based modeling and simulation (ABMS).
Twinner can represent the complexity of multi-level dynamic social systems and enables stakeholders to test their theories and strategies for addressing their societal mission.
Twinner has three main components (see Figure \ref{fig:SDT_Architecture}): data infrastructure, simulation engine and UI.

\subsection{Data Infrastructure}
\label{sec:Data-Infrastructure}

Data infrastructure integrates and securely stores a variety of multi-source multi-granularity multi-modal \textit{real-world} data (i.e., verifiable data directly collected from existing trusted third-party data sources) and \textit{synthetic} data (i.e., generated to mimic the characteristics and distributions of real-world data without exposing sensitive information). 
This information is accessible only to authorized users, ensuring data \textit{privacy} and \textit{security}.
We consider employing infrastructure data (e.g., residential infrastructure, public spaces and facilities, public and government buildings, educational infrastructure, industrial infrastructure, commercial infrastructure, transportation infrastructure, healthcare infrastructure, telecommunications infrastructure) and GIS data as real-world data, and population data and household data as synthetic data.

For the time being, we employed the real-world and synthetic data related to Norway as described below. 

\subsubsection{Norwegian Residential Infrastructure Data}
\label{sec:Norwegian-School-and-Residence-Data}

These real-world infrastructure data include georeferenced data about residential buildings in Norway. 
The data on residential buildings are gathered from two publicly available data subsets from Matrikkelen. 
Matrikkelen (the Norwegian Cadastre is the official Norwegian property register, run by the Norwegian Mapping Authority, and it contains georeferenced information, including property boundaries, areas, buildings, residences, and addresses. 
The two subsets that were used in the Social Digital Twinner were on apartment-level data and a dataset that included building type (e.g., studio apartment, detached house, row house, apartment complex in different sizes, cabin or garage)\footnote{Matrikkelen - adresse leilighetsnivå.
\url{https://kartkatalog.geonorge.no/metadata/matrikkelen-adresse-leilighetsnivaa/365b0591-b536-42a6-a20d-22e404fbfe55}, accessed on 13/03/2025.}\textsuperscript{,}\footnote{Matrikkelen - bygningspunkt. \url{https://kartkatalog.geonorge.no/metadata/matrikkelen-bygningspunkt/24d7e9d1-87f6-45a0-b38e-3447f8d7f9a1}, accessed on 13/03/2025.}. 
Identifying building types was crucial to accurately place the population and prevent non-residential structures, such as garages, from being treated as residential buildings. 
Additionally, determining apartment levels ensured the correct number of residential units, preventing scenarios where an entire apartment complex would mistakenly be allocated to a single family.

\subsubsection{Norwegian Educational Infrastructure Data}

These real-world infrastructure data include georeferenced data about compulsory schools and high schools in Norway.
The point data, addresses, and contact information for compulsory schools and high schools are sourced from the National School Register (Nasjonalt Skoleregister), managed by the Norwegian Directorate for Education and Training (Utdanningsdirektoratet, Udir), and retrieved via GeoNorge\footnote{ Geonorge grunnskoler. \url{https://kartkatalog.geonorge.no/metadata/grunnskoler/db4b872f-264d-434c-9574-57232f1e90d2}, accessed on 13/03/2025.}\textsuperscript{,}\footnote{Geonorge VGS. \url{https://kartkatalog.geonorge.no/metadata/videregaaende-skoler/c8acfd4f-c285-45a6-9a9b-3ab8d7d3af19}, accessed on 13/03/2025.}. 
Information on the number of pupils per grade at each school is sourced from \textit{Grunnskolens Informasjonssystem} (GSI), a system for registering information about primary and lower secondary education in Norway, and is managed by Udir\footnote{Grunnskolens informasjonssystem (gsi). \url{https://gsi.udir.no/informasjon/}, accessed on 13/03/2025.}. 
Information on the number of students per grade in each high school is also obtained from one of Udir's data systems; \textit{Elevtall i videregående skole}\footnote{Utdanningsdirektoratet: Elevtall i videregående skole – fylker og skoler.
\url{https://www.udir.no/tall-og-forskning/statistikk/statistikk-videregaende-skole/elevtall-i-videregaende-skole/elevtall-fylker-og-skoler/}, accessed on
13/03/2025.}. 

\subsubsection{Synthetic Norwegian Population Data}
% New suggested text (VAFA) date:250130
The synthetic Norwegian population is a full-scale demographically fit replica generated from high-featured and high-quality microdata sources like EU-SILC living condition data\footnote{EU-SILC Living Condition Data,
\url{https://ec.europa.eu/eurostat/web/microdata/european-union-statistics-on-
income-and-living-conditions}, accessed on 13/03/2025.}\textsuperscript{,}\footnote{Gesis EU-SILC, \url{https://www.gesis.org/en/missy/metadata/EU-
SILC/}, accessed on 13/03/2025.}. 
The synthetic Norwegian population based on EU-SILC represent all adults from sixteen and up in all European countries connected to the EU-SILC initiative. 
EU-SILC data are census data standardised across countries and comprise a large number of variables of interest to social simulations.
EU-SILC living condition data comprise a wide range of variables, from social and economic status and workforce connections to self-perceived health and confidence in various parts of society.
Combining this population with aggregated data from the local statistical agencies, we can produce households with kids at various national levels, like municipalities (the household synthesiser). 
Microdata from other high-quality sources, like Youth Data for Norway\footnote{Ungdata (2024), \url{https://www.ungdata.no/}, accessed on 13/03/2025.}, can be used to build a kids' population for households. 
Through demographic keys, the synthetic population of adults can be combined with other high-quality microdata sources like medical data\footnote{Hunt databank, \url{https://www.ntnu.edu/hunt/databank}, accessed on 13/03/2025..}\textsuperscript{,}\footnote{The tromsø study, \url{https://uit.no/research/tromsostudy}, accessed on 13/03/2025..} or municipality data to strengthen the applicability further.
The current population synthesiser generates synthetic populations by advanced generative machine learning (ML) methods trained on weighted EU-SILC data \cite{Falck2024}. 
A full-scale replica of a municipality's population with households and children facilitates studies of intergenerational dynamics and social behaviour linked to, i.e. youths' exclusion from school and work (NEET).

\subsection{Simulation Engine}

%%% The following command should be issued somewhere in the first column of the final page of your paper.

%\balance

\begin{figure}[h]
  \centering
  \includegraphics[width=0.99\linewidth]{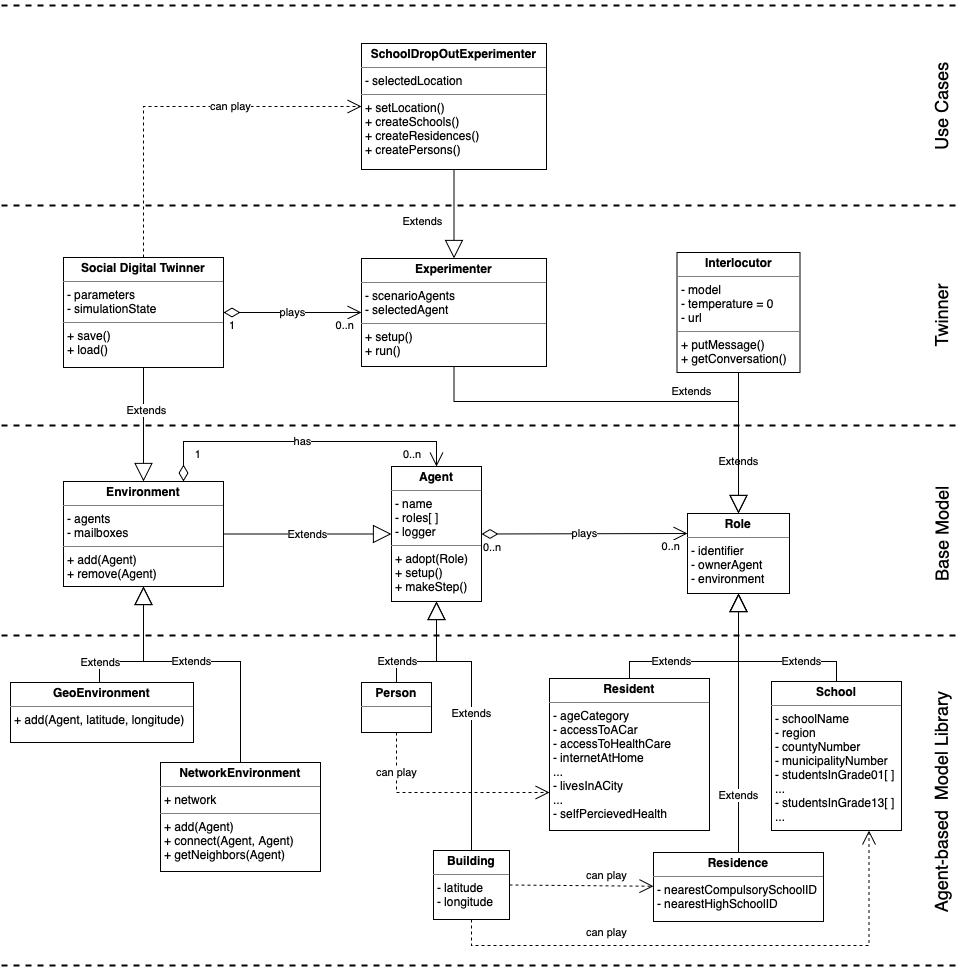}
  \caption{Organizational Agent-based Model Library of Social Digital Twinner}
  \label{fig:SDT_SimulationEngine}
\end{figure}

The simulation engine is a versatile and powerful agent-based simulation tool designed to support the study and analysis of complex social phenomena. 
It enables the simulation of diverse societal processes with dynamic interactions. 
The simulation engine is composed of four layers as shown in Figure \ref{fig:SDT_SimulationEngine}.

The \textit{base model} layer uses the Agent/Group/Role (AGR) organizational multi-agent modeling approach \cite{Ferber2004,Roussille2022} and facilitates the integration of the other layers. 
It contains three basic concepts: Agent, Environment (i.e. group) and Role. 
Agents are dynamic, communicative entities that play roles within groups.
Environments function as identifiers for particular contexts (physical or logical) where specific roles are jointly performed by groups of agents, effectively bringing together agents involved in collaborative efforts. Agents are permitted to communicate only if they belong to the same environment.
Roles are activity patterns performed by agents within an environment.

The \textit{agent-based model library} layer comes with predefined environments, roles, and agents dedicated to the SDT domain.
For the time being, as environments, we have \textit{GeoEnvironment} for holding agent with geographical positions and \textit{NetworkEnvironment} for connecting the agents to each other.
As agents, we have the \textit{Person} and \textit{Building} agents.
While the Person agents can play the \textit{Resident} role, the building agents can play \textit{School} and \textit{Residence} roles. 
The data specific to each role are coming from the data infrastructure: the Resident role uses the synthetic population data, the School role uses the educational infrastructure data and the Residence role uses the residential infrastructure data.

The \textit{twinner} layer comes with the \textit{Social Digital Twinner} agent, the \textit{Experimenter} role and the \textit{Interlocutor} role.
Social Digital Twinner is the environment and the agent who brings together the other agents and is responsible for setting up and running the simulation experiments specified as Experimenter roles. 
The Interlocutor role allows all agent to have LLM-based conversation capabilities by using the agent's properties, a log of past actions and the conversation history.
The LLM then generates responses that are aligned with the agent's behavior.
To minimize the influence of internal biases within the LLM, we set the temperature parameter's value to 0 and added the following system prompt at the beginning of the conversation:

\begin{quote}
    Be as concise as possible in your answers. If you do not know the answer to a question, simply answer as `I do not know.'.
\end{quote}

Then depending on the agent and the roles it plays, the following dynamically generated prompt is added to the conversation:

\begin{quote}
    As \texttt{\{ownerAgent.name\}}, your current role is \texttt{\{role.name\}} within the environment \texttt{\{environment\}}. You have access to \texttt{\{role\_specific\_data\}}. You are empowered to act directly as \texttt{\{role.name\}}. Analyze the user's question or request, leverage your knowledge and the provided data, and generate the best possible answer or solution. If you need additional reasoning steps, please outline them clearly. Then, finalize your response as \texttt{\{role.name\}}.
\end{quote}

The \textit{use cases} layer comes with predefined use cases specified as roles that Social Digital Twinner can play.
For the time being, it contains the \textit{SchoolDropOutExperimenter} role which specifies a use case focusing on a scenario of school dropouts (see Section \ref{sec:Case-Study-School-Dropout-of-NEETs}).

%There are also predefined use cases using the aforementioned organizational abstractions, allowing for rapid development and customization of simulation scenarios.

\subsection{User Interface (UI)}

\begin{figure}[h]
  \centering
  \includegraphics[width=1.0\linewidth]{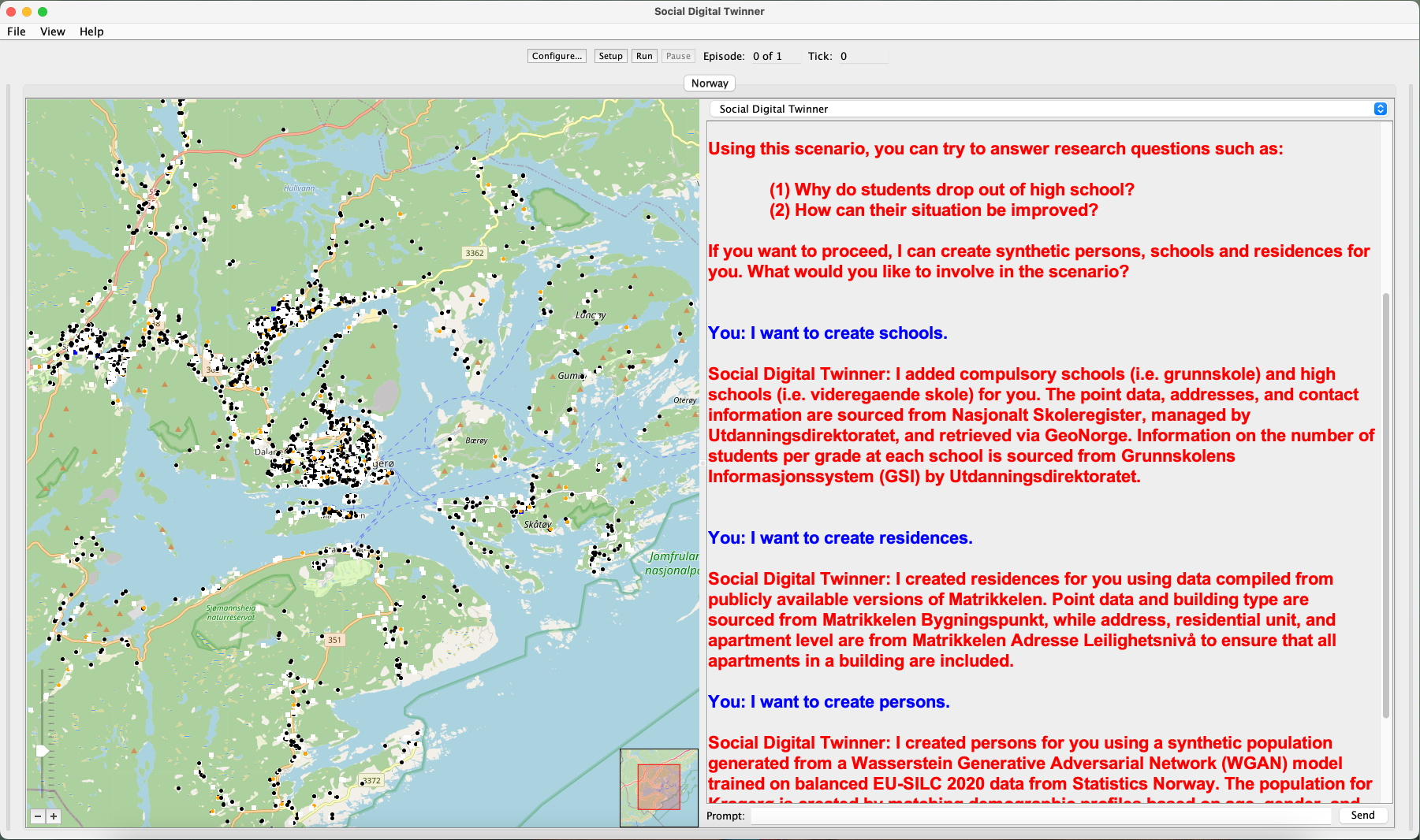}
  \caption{The UI of Social Digital Twin of Norway}
  \label{fig:STDN_UI}
\end{figure}

The UI of Twinner combines a visual map-based representation of agents  with a text-based interaction window, enabling real-time communication and querying of artificial agents by users (Figure~\ref{fig:STDN_UI}).

On the left side of the UI, a map visualization panel provides a detailed geospatial view of an urban environment, complete with roads, buildings, and topographical features. Agents within the simulation are represented as colored dots or rectangles (or other shapes) scattered across the map, indicating their presence in various locations. 
The selected agent is highlighted in red to distinguish it from the other agents. 
This visual representation allows users to intuitively locate and select agents for interaction, facilitating exploration of the simulated environment.

On the right side, the text-based interaction window serves as the primary medium for engaging with agents. 
The UI supports a structured dialogue format, where user queries appear in blue text under the label “You:”, and selected agent's responses are shown in red under the label as selected agent name\footnote{Note that this user interface is strictly between the user and the selected agent.}.
The dialogue can progress seamlessly with follow-up questions, demonstrating the LLM-augmented agent’s ability to maintain context and deliver coherent, informative answers.
At any time, the user can select another agent and engage in a dialogue with it.

%%%%%%%%%%%%%%%%%%%%%%%%%%%%%%%%%%%%%%%%%%%%%%%%%%%%%%%%%%%%%%%%%%%%%%%%

\section{Case Study: School Dropout of NEETs  in Kragerø}
\label{sec:Case-Study-School-Dropout-of-NEETs}

We illustrate a use case focusing on a scenario of school dropouts among the phenomenon of NEET (Not in Education, Employment, or Training) \cite{Oydgard2024} in Kragerø, Norway. 
The scenario involves schools, residences, local citizens, and households.
Students are expected to attend their nearest compulsory schools or high schools within their neighborhood. 
However, for those in rural areas without a nearby high school, commuting to an urban high school is necessary.  
If students miss more than 10 consecutive school days, they are classified as dropouts. 
For demonstration purposes, all students in rural areas drop out, showcasing how the Twinner can monitor and represent such societal issues.
%A video recording of this demonstration is available via the link provided in the footnote.\footnote{\url{https://drive.google.com/file/d/1Ju6vZHXk7djYxszjHkag6VqWpAufObXn/view?usp=share\_link}. If the paper is  accepted, we will upload an enhanced recording to a publicly accessible platform such as YouTube.}.

Traditional social simulation platforms would require scripting specific dropout conditions and manually configuring scenario data. 
In contrast, our platform allows users to formulate hypotheses, inspect agent responses, and modify behavior on the fly using natural language (Figure~\ref{fig:STDN_UI}). 
For example, a user can directly ask an individual student agent why they dropped out and receive an answer grounded in the agent's synthetic profile and local context. This enables a more exploratory and iterative policy development process, something not easily achievable in most pre-scripted ABM tools.

%%%%%%%%%%%%%%%%%%%%%%%%%%%%%%%%%%%%%%%%%%%%%%%%%%%%%%%%%%%%%%%%%%%%%%%%

\section{Discussion}
\label{sec:Discussion}

In this section, we explore key aspects of our study, including privacy and security considerations (Section \ref{sec:Privacy-and-Security}), the role of democratization in access and usage (Section \ref{sec:Democratisation}), and the implications for human behavior simulation (Section \ref{sec:Human-Behavior-Simulation}).

\subsection{Privacy and Security}
\label{sec:Privacy-and-Security}

Since Twinner creates digital replicas of real buildings and schools, concerns may arise regarding potential privacy issues for people within these spaces.
First of all, the privacy of the personal data has already been considered when retrieving georeferenced data.
As mentioned in Section \ref{sec:Norwegian-School-and-Residence-Data}, the school and residence data used are based on publicly available maps from the Norwegian Mapping Authority, rather than restricted versions that contain personal data.

Secondly, the synthetic population is anonymised by replicating original and anonymised EU-SILC census microdata and upscaling from replicas to full-level municipality using aggregated national statistics\footnote{Norwegian statistical agency (2024), \url{https://www.ssb.no/}, accessed on 13/03/2025.} to fit demographic profiles. 
The population generation and validation procedures are described in \cite{Falck2024}. 
Even with the risks of de-anonymising (i.e. re-identification) through reverse engineering \cite{Wong2024,Torres2024}, these data are still based on training anonymised but restricted-use data.
Since Twinner is allocating randomly the synthetic population to residences in a municipality, any similarities between people living at the address and in the simulation are incidental and not resulting from a privacy breach.  
We are aware that such pseudo-privacy breaches might evoke stakeholders' concerns.

Concerning security, as discussed in Section 3.1, the data used in Twinner is securely accessed, mitigating the risk of unauthorized breaches. 
Twinner implements robust cybersecurity measures (e.g., access control and authentication) to protect digital twins and associated data, ensuring that only authorized entities can access sensitive information. 
By adhering to strict security protocols, Twinner minimizes privacy risks and prevents potential misuse of data.

\subsection{Democratisation}
\label{sec:Democratisation}

The Twinner platform advances the democratization of societal digital twins by leveraging agent-based modeling (ABM) frameworks, as proposed in \cite{Birks2020}. 
By enabling stakeholders like policymakers and researchers to create on-demand, interactive social digital twins via natural language interfaces, the platform lowers technical barriers, empowering non-specialists to simulate complex human behaviors and their interdependencies with physical infrastructure—a critical gap highlighted in current urban digital twin projects (e.g., Rotterdam, Singapore). 

This democratization aligns with the vision of integrating synthetic populations into city-scale models to explore emergent social outcomes, such as how transport policy changes might influence crime patterns or resource allocation. 
By abstracting the complexities of ABM and real-time data assimilation—key challenges noted in the paper—the platform makes advanced computational laboratories accessible, fostering inclusive, cross-sector collaboration to address "wicked problems" like inequality. 

However, it also inherits ethical dilemmas discussed in the literature, including biases in social data, privacy risks from granular behavioral modeling, and transparency in uncertainty-laden simulations. Addressing these requires governance frameworks co-developed with interdisciplinary experts and communities, ensuring democratization does not compromise equity. 
Ultimately, the platform bridges the paper’s call for holistic, human-centric digital twins while expanding participatory decision-making in socio-physical systems.

\subsection{Natural Language Interaction}
\label{sec:Human-Behavior-Simulation}

Our LLM-enhanced agents can provide direct, fact-based answers regarding their internal properties using natural language\footnote{It should be noted that, in the current implementation, the LLM is used exclusively as a natural language interface for human users to communicate with artificial agents. While the architecture does not restrict agents from using LLMs to communicate with each other, all agent-to-agent interactions presently occur via conventional multi-agent system (MAS) mechanisms, such as direct message passing or event-based coordination. The LLM serves solely as a user-facing conversational layer, facilitating intuitive access to agent data and behaviors.}. 
Their responses are grounded explicitly in their underlying data, which ensures both transparency and verifiability. For example, building agents representing schools can report specific details, such as the number of students in each grade.
In addition, these agents are capable of aggregating and summarizing their properties. For instance, building agents playing the role of schools can readily provide aggregated data, such as the total number of students enrolled between Grade 2 and Grade 5. Such aggregated information remains straightforward to verify.
Furthermore, the agents can generate higher-level insights by aggregating and summarizing information beyond simple property values. For example, agents representing residents can suggest specific improvements to their living conditions, potentially preventing them from dropping out of school.

Simulating realistic human behavior remains a fundamental yet complex challenge. While LLMs can produce highly convincing, human-like outputs and significantly enhance our understanding of human behaviors, we acknowledge that they cannot fully replace humans at present. Instead, they serve as powerful supplementary resources, complementing human insight and analysis \cite{Harding2024,Dillion2023,Frank2023}.

\section{Conclusion and Future Work}
\label{sec:Conclusion-and-Future-Work}

We present the Social Digital Twinner, an advanced and democratized platform designed to build on-demand social digital twins for delivering deep insights into societal challenges at various scales, enhancing decision-making and policy development with greater precision and reliability.
Twinner achieves greater realism, flexibility, and scalability thanks to the combination of cutting technologies like ABM, ML and LLMs.

%One of the main challenges in developing STDN was integrating diverse data sources to create a cohesive, accurate and privacy-preserving \cite{Milosevic2024} model.
%This requires advanced technologies such as IoT, AI, and big data analytics.

%As the platform advances in sophistication, it will be crucial to manage its growing complexity and ensure its scalability. 
%
%Consequently, as a future work, we plan to demonstrate the tool’s capabilities against an increasing number of societal use cases, showcasing its potential to democratize dynamic policy experimentation through our easily accessible tool and fostering informed decision-making.

While the Social Digital Twinner introduces an innovative and flexible approach to social digital twinning, the techniques and methodologies employed in our framework have not yet been formally evaluated. 
A thorough assessment of the system's accuracy, usability, and effectiveness in real-world decision-making scenarios is essential to validate its practical impact. 

As part of our future work, we plan to conduct systematic evaluations of the employed techniques through empirical studies, benchmarking experiments, and stakeholder engagement. 
This will enable us to refine the platform’s capabilities, ensure robustness, and further enhance its role in facilitating data-driven societal decision-making.

%Additionally, other countries may also seek to adopt the SDTN infrastructure and establish connections with it. 
%In this context, it is likely that SDTN will evolve into a federated digital twin, enabling collaboration and shared learning across multiple interconnected systems \cite{Vergara2023}.

%%%%%%%%%%%%%%%%%%%%%%%%%%%%%%%%%%%%%%%%%%%%%%%%%%%%%%%%%%%%%%%%%%%%%%%%

\section*{Acknowledgment}

The work reported here is part of the EU Horizon URBANE project\footnote{URBANE, \url{https://www.urbane-horizoneurope.eu}, last access on May 9, 2025.} \cite{Franklin2025} with grant agreement No. 101069782.

%
% ---- Bibliography ----
%
% BibTeX users should specify bibliography style 'splncs04'.
% References will then be sorted and formatted in the correct style.
%
\bibliographystyle{splncs04}
\bibliography{references}
\end{document}